\documentstyle[prl,aps,amssymb,amsmath,multicol,epsf]{revtex}

\begin{document}

\title{Distributed Entanglement as a Probe for the Quantum Structure
 of Spacetime} 
 \author{Pieter Kok\cite{pieter}$^1$, Ulvi Yurtsever$^1$, Samuel L.\
 Braunstein$^2$, and Jonathan P.\ Dowling$^1$} 
\address{$^1$ Jet Propulsion Laboratory, California Institute of
 Technology, Quantum Computing Technologies Group, \\ Mail Stop
 126-347, 4800 Oak Grove Drive, Pasadena, California 91109} 
\address{$^2$ Informatics, Bangor University, Bangor LL57 1UT, UK}

\maketitle
\begin{abstract}

Simultaneity is a well-defined notion in special
relativity once a Minkowski metric
structure is fixed on the spacetime continuum (manifold) of events.
In quantum gravity, however, the metric is not expected to be a fixed,
classical structure, but a fluctuating quantum operator which
may assume a coherent superposition
of two classically-distinguishable values. A natural question to ask
is what happens to the notion of simultaneity and
synchronization when the metric is in a quantum superposition.
Here we show that the resource of distributed entanglement of the
same kind as used by Jozsa {\em et al.} [Phys.\ Rev.\ Lett.\ {\bf 85},
2010 (2000)] gives rise to an experimental probe
that is sensitive to coherent quantum fluctuations
in the spacetime metric.

\end{abstract}

\medskip


\medskip

\begin{multicols}{2}

For a given choice of a Minkowski metric structure on the spacetime
continuum, simultaneity is a uniquely defined notion in special
relativity. Although there is an infinite class of distinguishable but
equivalent Minkowski metrics on the spacetime manifold, the specific
metric that is to be used is a philosophical problem that seems  to
have no consequences for real experiments. In particular, it does not
have any computational implications for classical physics
\cite{reichenbach,ellis,grunbaum,malament,redhead,anderson}. In quantum
gravity, however, the metric is not expected to be a fixed, classical
structure, but a fluctuating quantum operator. In particular, it is
conceivable that the metric can be in a coherent superposition of two
classically-distinguishable values. Experimentally, there exist well
known protocols to construct the classical metric once a labeling of
actual events as spacetime points is carried out. One particular such
protocol involves the synchronization of the clocks of two distant
observers (Alice and Bob) at rest with respect to each
other. Recently, clock  synchronisation received renewed interest with
the added resource of shared entanglement between Alice and Bob
\cite{jozsa}. A natural question to ask is what happens to the notion
of simultaneity and synchronization when the metric is in a quantum
superposition. Here we show that the resource of distributed
entanglement of the same kind as used in \cite{jozsa} gives rise to an
experimental probe that is sensitive to coherent quantum fluctuations
in the spacetime metric.

In special relativity, given a fixed Minkowski metric $g$ on spacetime
${\bf R}^4$, simultaneity is defined as follows: let $u^{\alpha}$ be
the four-vector of an inertial observer, Alice, and let $P$ be an
event along the world-line of Alice. Then Alice's surface of
simultaneity at $P$ is the set of all events $Q$ such that the
space-like vector (or geodesic) $S^{\alpha}$ joining $P$ to $Q$ is
orthogonal to $u^{\alpha}$: $g_{\alpha \beta}u^{\alpha}
S^{\beta}=0$. This definition is formulated entirely in terms of
physically observable quantities, and, given a fixed metric $g_{\alpha
  \beta}$, is implemented in practice using the Einstein
synchronization protocol. 

The protocol works as follows: suppose Alice and Bob are separated by
a (large) distance $d$.  Alice sends a light signal to Bob, who uses a
mirror to return the signal immediately. Alice then measures the time
interval between the departure at $t_1$ and the arrival at $t_3$ of
the signal: ($t_3 - t_1$), and defines the half-way time $t_2$ through
this interval as 
\begin{equation}
 t_2 \equiv t_1 + \frac{1}{2} \left( t_3 - t_1 \right)\; .
\end{equation}
By construction, the spacelike vector joining the event at time $t_2$
on Alice's worldline to the event of reflection $t_2'$ in Bob's mirror is
orthogonal to Alice's four velocity. In other words, $t_2$ and $t_2'$
lie on a surface of simultaneity for Alice (and Bob), according to the
above definition. Alice now tells Bob that the time of reflection
(which Bob recorded, e.g., by measuring the impulse on the mirror) was
at $t_2$ on her clock. Bob can then adjust his clock so that his
measured time at this event coincides with $t_2$, and we have
therefore obtained clock synchronisation in accordance with the above
definition.  

\begin{figure}[t]
 \begin{center}
  \epsfxsize=8in
  \epsfbox[-10 10 800 200]{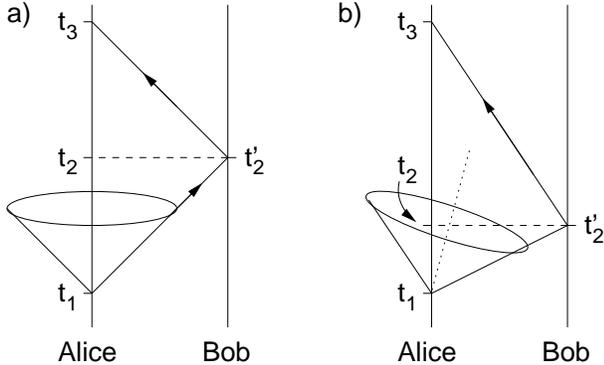}
 \end{center}
 \caption{Einstein synchronization 
        using the round-trip travel time of a light signal. The
        location of the reflection event $t_2'$ on Bob's worldline
        which is defined to be simultaneous with Alice's $t_2$,
        depends entirely on the given Minkowski metric $g_{\alpha
        \beta}$ on spacetime: a) the synchronization process for the
        "standard" metric (only the light cones of $g_{\alpha \beta}$
        are shown), b) Einstein synchronization with a different
        Minkowski metric.}
 \label{fig1}
\end{figure}

Notice that this protocol depends crucially on the specific Minkowski
metric $g_{\alpha \beta}$ that is fixed from the outset (see Fig.\
\ref{fig1}). In general, there exists an infinite class of distinct
Minkowski metrics on the manifold ${\bf R}^4$: For any diffeomorphism
$\phi : {\bf R}^4 \longrightarrow {\bf R}^4$, the metric $g' \equiv
{\phi}^{\ast}(g)$ is an element in that class
(where ${\phi}^{\ast}$ denotes the tensorial ``pullback" map
associated to the diffeomerphism $\phi$) distinct from $g$ unless
$\phi$ happens to be a transformation in the Poincare group (i.e.\ a
Lorentz transformation combined with translations). Which specific
metric represents the real Lorentz structure is an operational
question that can in principle be answered by experiment;
nevertheless, since physics is invariant under isometries, these
different Minkowski  metrics are in any case physically equivalent to
each other \cite{reichenbach,ellis,grunbaum,malament,redhead,anderson}.

So far, the discussion has been purely classical. In quantum mechanics 
Alice and Bob might share entanglement, and the spacetime metric is
generally no longer a fixed background structure, but is subject to
quantum fluctuations. In this paper, we will show how this extra
resource of shared distributed entanglement can be used as an
experimental probethat is sensitive to coherent quantum fluctuations
in the spacetime metric. 

Consider our two observers, Alice and Bob, who initially are
co-located and share a singlet state of two qubits whose computational
basis states $|0\rangle$ and $|1\rangle$ correspond to nondegenerate
(distinct) energy levels $E_0$ and $E_1$ (where we define without loss
of generality $E_1 > E_0$). The initial quantum state of the joint
system is given by
\begin{equation}
|\Psi\rangle = \frac{1}{\sqrt{2}}\, (\, |0 \rangle_A  |1 \rangle_B -
| 1 \rangle_A  | 0 \rangle_B ) \; .
\end{equation}
Throughout this paper, the subscripts $A$ and $B$ Alice and Bob
respectively. Suppose this entanglement is now distributed by letting
Alice move a large distance $d$ away from Bob. After the distribution,
when Bob and Alice are at relative rest again, the state of the system
can be written in the form
\begin{eqnarray}
|\Psi\rangle & = & \frac{1}{\sqrt{2}}\, \left( \, e^{-i \Omega_0 {\tau_A}}
|0 \rangle_A \otimes e^{-i \Omega_1{\tau_B}} |1 \rangle_B \right. \nonumber \\
&& \qquad - \left.
e^{-i \Omega_1{\tau_A}}| 1 \rangle_A  \otimes
e^{-i \Omega_0{\tau_B}}| 0 \rangle_B \right) \; ,
\end{eqnarray}
where $\tau_A$ and $\tau_B$ are the proper times that elapsed in Alice
and Bob's frame during the entanglement transport,
and $\hbar \Omega_0$ and $\hbar \Omega_1$ are the
ground and excited state energies, respectively. Up to an overall phase,
the state Eq.\,(3) can be rewritten as
\begin{equation}
|\Psi\rangle = \frac{1}{\sqrt{2}}\, \left( \, |0 \rangle_A  |1 \rangle_B -
e^{i \Omega{(\tau_B - \tau_A )}}| 1 \rangle_A  | 0 \rangle_B \right) \; ,
\end{equation}
where $\Omega \equiv \Omega_1 - \Omega_0 $.
When Alice and Bob are at relative rest (comoving: $\tau_A
=\tau_B$), $\Psi$ is a dark
state since its time evolution corresponds to multiplication
by an overall phase factor.

Alice and Bob now execute the clock synchronization protocol
introduced by Jozsa {\em et al}.\ \cite{jozsa}. First Alice makes a
measurement on her qubit in the $\{ |\pm\rangle_A \}$ basis:
\begin{eqnarray}
| + \rangle_A & \equiv & \frac{1}{\sqrt{2}} (|0 \rangle_A + | 1 \rangle_A )\; ,
\nonumber \\
| - \rangle_A & \equiv & \frac{1}{\sqrt{2}} (|0 \rangle_A - | 1 \rangle_A )\; ,
\end{eqnarray}
and communicates the result classically to Bob. Assume that Alice
obtains the result $| + \rangle_A$. Bob then knows that his
qubit is in the reduced state
\begin{equation}
|\phi^{(-)}\rangle_B = \frac{1}{\sqrt{2}}\, \left( \, |1 \rangle_B -
e^{i \Omega{(\tau_B - \tau_A )}} | 0 \rangle_B \right) \; ,
\end{equation}
obtained via the projection $P_{|+ \rangle_A } \Psi$ from
the state Eq.\,(4). This is the same protocol used in the clock
synchronization application of \cite{jozsa}, and the phase
$\Omega (\tau_B - \tau_A )$ is the ``Preskill" phase which makes this
application difficult to implement in practice\cite{ulvi00}. Here we are
not interested in a synchronization protocol, however, and the important
thing about the state Eq.\,(6) is that it is a pure state (though an
unkown pure state) as Bob can verify experimentally.

Suppose now that the background spacetime on which the above
protocol is implemented has a metric subject to quantum fluctuations.
Let us assume that the quantum state of the metric $|g\rangle$ is in a coherent
superposition of two macroscopically-distinguishable (orthogonal) states
$|g_0 \rangle$ and $| g_1 \rangle$ given by
\begin{equation}
|g\rangle = \alpha |g_0 \rangle + \beta | g_1 \rangle \; ,
\end{equation}
where $\alpha$ and $\beta$ are complex numbers with
$|\alpha |^2 + |\beta |^2 = 1$. Assume, furthermore, that the proper
time elapsed during Alice's trip to her final destination differs for
the two metrics $g_0$ and $g_1$ by a (small) time interval $\Delta$.
Under these assumptions, after entanglement distribution
the total state of the singlet system plus the metric can be written
in the form [compare Eq.\,(3)]
\[
|\Psi ,g\rangle = \;\;\;\;\;\;\;\;\;\;\;\;\;\;\;
\;\;\;\;\;\;\;\;\;\;\;\;\;\;\;\;\;\;\;\;\;\;\;\;\;\;\;\;\;\;
\;\;\;\;\;\;\;\;\;\;\;\;\;\;\;
\]
\begin{eqnarray}
\frac{1}{\sqrt{2}}  \left[
| 0 \rangle_A  e^{-i \Omega_1{\tau_B}} |1 \rangle_B \otimes 
\left(
e^{-i\Omega_0 \tau_A} \alpha |g_0 \rangle
 + e^{-i \Omega_0 (\tau_A + \Delta )} 
\beta |g_1 \rangle \right) \right. \nonumber \\
 - \left.
|1 \rangle_A e^{-i \Omega_0{\tau_B}} |0 \rangle_B  \otimes  \left(
e^{-i\Omega_1 \tau_A} \alpha |g_0 \rangle + e^{-i \Omega_1 (\tau_A + \Delta )}
\beta |g_1 \rangle \right) \right] \; . \nonumber
\end{eqnarray}
\begin{equation}
\;
\end{equation}
From the point of view of Alice and Bob, the quantum state of the
gravitational field is inaccessible via any direct observation; therefore, their
joint state is described by tracing over the gravitational part of
the wave function Eq.\,(8):
\begin{eqnarray}
\rho_{AB} & = & {\rm Tr}_g \left[ | \Psi, g\rangle \langle \Psi, g | \right] 
\nonumber \\
& = & \frac{1}{2} \left[ \; |0\rangle_A \langle 0|_A \otimes |1\rangle_B
\langle 1|_B
+ |1\rangle_A \langle 1|_A \otimes |0\rangle_B \langle 0|_B \right. \nonumber \\
& - & \overline{W} \; |0\rangle_A \langle 1|_A \otimes |1\rangle_B
\langle 0|_B
-  W \; |1\rangle_A \langle 0|_A \otimes |0\rangle_B \langle 1|_B 
\left. \right] \;,
\end{eqnarray}
where $W$ denotes the complex number
\begin{equation}
W \equiv e^{i\Omega (\tau_B - \tau_A )}
\left( |\alpha|^2 + e^{-i\Omega \Delta}|\beta|^2 \right) \; .
\end{equation}
What will happen when Bob and Alice carry out the protocol described
above [Eqs.\,(5-6)]? Alice performs the measurement in her $\{
|\pm\rangle_A \}$ basis, and obtains a random string of outcomes
$\{``+",``-"\}$. She sends this bit string to Bob. In practice, this
means that Alice tells Bob which of her enumerated singlet-halfs from
a large ensemble are projected onto the $|+\rangle_A$ state. Note that
this calculation describes an experiment in which the various
ensemples are obtained via repeated application of the entanglement
distribution and measurement protocols described above, in each such
application the metric being in the same coherent state described by
Eq.\,(7).] The state $\rho_{AB}$ in Eq.\,(9) then collapses to the
density matrix 
\begin{eqnarray}
\rho_{AB} & \mapsto & \frac{P_{|+\rangle_A} \rho_{AB} P_{|+\rangle_A}}
{{\rm Tr} \left( P_{|+\rangle_A} \rho_{AB} P_{|+\rangle_A} \right)}
\nonumber \\
& = & 
\frac{1}{2} |+\rangle_A \langle+|_A   \nonumber \\
& \otimes& \left[ \;  |0\rangle_B \langle 0|_B
+ |1\rangle_B \langle 1|_B 
- W \; |0\rangle_B \langle 1|_B
- \overline{W} \; |1\rangle_B \langle 0|_B  \right] \; . \nonumber
\end{eqnarray}
\begin{equation}
\;
\end{equation}
Bob's reduced state, therefore, is given by
\begin{eqnarray}
\rho_B & = & \frac{1}{2} \left[ \;  |0\rangle_B \langle 0|_B
+ |1\rangle_B \langle 1|_B \right. \nonumber \\
& - & \left . W \; |0\rangle_B \langle 1|_B
- \overline{W} \; |1\rangle_B \langle 0|_B  \right] \; .
\end{eqnarray}
Contrast Eq.\,(12) with the pure state Eq.\,(6). The state Eq.\,(12) is
pure if and only if
\begin{equation}
{\rm Tr} \; {\rho}^2 = \frac{1}{2} (1+|W|^2) = 1 \; ,
\end{equation}
which is possible if and only if $|W|=1$. But
\begin{eqnarray}
|W|^2 & = & 1 - 4 |\alpha|^2 |\beta |^2 \sin^2 \left( \frac{\Omega
\Delta}{2} \right)
\nonumber \\
& \approx & 1 - |\alpha|^2 |\beta |^2 \Omega^2 \Delta^2 \; ,
\end{eqnarray}
where the approximate equality assumes $\Omega \Delta \ll 1$. In
general, Bob's state at the end of the protocol is mixed, and if
experiment can distinguish this ``decoherence" effect from other
sources of decoherence, it provides a possible probe for the quantum
fluctuations in the spacetime metric.

A similar gravitational decoherence effect could be produced by using
a single clock qubit carried by Alice through the region with the
fluctuating metric. However, the advantages of using the singlet state
as a probe are twofold: First, the singlet state is immune to phase
decoherence effects that interact with both qubits in the same way; so
it provides a more localized probe than a single qubit would. Second,
the singlet state allows the final measurement process to be performed
in a region (Bob's location) arbitrarily distant from the region where
the gravitational interaction takes place (Alice's worldline); thus
there is no danger of the final measurement process ``collapsing" the
state of the metric leading to a false negative result (a pure-state
outcome for Bob).  

What are the prospects for this probe to be a realistic one? For Planck
scale fluctuations in the metric, for which $\Delta \sim 10^{-43}$sec
(the Planck time), an observable effect will be present if $\Omega
 \sim 10^{43} $Hz, which corresponds, not surprisingly,
to energies $\hbar \Omega$ of the order of the rest energy corresponding to the
Planck mass (roughly the mass of a grain of sand). This is a macroscopic
amount of energy (difference): the entangled state Eq.\,(2) would
have to be a macroscopic, ``Schr\"odinger's cat" state and
preserve its coherence over the large distance $d$. Since many
non-gravitational sources of decoherence would tend to destroy such
states rather quickly, the prospects for a succesful experiment to
probe for Planck-scale geometry fluctuations are not good.

There are, however, recent intriguing suggestions
that microscopic black holes could be produced in large hadron colliders
currently under construction \cite{string}. According to these suggestions, string
theory may present a new length scale much larger than the Planck
length at which the gravitational interaction becomes a dominant force,
allowing much lower energy thresholds for black hole production. The
black holes produced in future hadron colliders may be as light as 1TeV in
mass, correponding to a length scale
on the order of $10^{-17}$cm. This new mass-length scale
involves the geometry of compactified
dimensions in string theory, but if these small black holes have
gravitational signatures in the non-compactified dimensions
that have length scales comparable to $10^{-17}$cm, or equivalently
time scales of $10^{-27}$sec, then our probe would be sensitive to such
fluctuations provided $\Omega \sim 10^{27}$Hz, which corresponds to
energies on the order of $\hbar \Omega \sim 1$TeV. This is a {\em
mesoscopic} energy scale, and it is not inconceivable that
entangled states of mesoscopic systems, like for example high
photon-number path-entangled states \cite{lee02,kok02}, can be produced
and maintained against non-gravitational decoherence just long enough
for the protocol we discussed above to be practical. We believe this
possibility is intriguing enough to deserve further study.

This work was carried out at the Jet Propulsion Laboratory, California
Institute of Technology, under a contract with the National
Aeronautics and Space Administration. 
In addition, P.K.\ would like to
thank William J.\ Munro for fruitful discussions, and the United
States National Research Council for financial support. Support was 
received from the Office of Naval Research, Advanced Research and
Development Activity, National Security Agency and the Defense
Advanced Research Projects Agency.

%
%
%

\end{multicols}
\end{document}